\documentclass[msom,nonblindrev]{informs3_hide}

\DoubleSpacedXI

\newcommand{\fa} {\ensuremath{\operatorname{FAIR-L}}\xspace}
\newcommand{\fb} {\ensuremath{\operatorname{FAIR-S}}\xspace}

\newcommand{\arr}{\ensuremath{\operatorname{Arr}}\xspace}
\newcommand{\serve}{\ensuremath{\operatorname{Serve}}\xspace}
\newcommand{\reserve}{\ensuremath{\operatorname{RESERVE}}\xspace}

\newcommand{\bmin}{{b}}

\newcommand{\fcfs}{\ensuremath{\operatorname{FCFS}}\xspace}
\newcommand{\nadaps}{\ensuremath{\operatorname{SAMP-S}}\xspace}
\newcommand{\nadap}{\ensuremath{\operatorname{SAMP}}\xspace}

\newcommand{\rank}{\ensuremath{\operatorname{Ranking}}\xspace}
\newcommand{\mgs}{\ensuremath{\operatorname{MGS}}\xspace}

\usepackage{xparse}
\NewDocumentEnvironment{myproof}{o}
  {\IfNoValueTF{#1}{\paragraph{{Proof.} }} {\paragraph{{#1} }} }
  {\hfill$\Halmos$}

\usepackage{natbib}
 \bibpunct[, ]{(}{)}{,}{a}{}{,}%
 \def\bibfont{\small}%
\TheoremsNumberedThrough     

\EquationsNumberedThrough    

\usepackage{macros_pan}

\SetAlFnt{\small}
\SetAlCapFnt{\small}
\SetAlCapNameFnt{\small}
\SetAlCapHSkip{0pt}
\IncMargin{-\parindent}

\usepackage{subcaption}
\usepackage{multirow}

\begin{document}




\TITLE{Group-level Fairness Maximization in Online Bipartite Matching}

\ARTICLEAUTHORS{%
\AUTHOR{Will Ma}
\AFF{Graduate School of Business, Columbia University, New York, NY 10027, \EMAIL{wm2428@gsb.columbia.edu}}
\AUTHOR{Pan Xu}
\AFF{Department of Computer Science, New Jersey Institute of Technology, Newark, NY 07102, \EMAIL{pxu@njit.edu}}
\AUTHOR{Yifan Xu}
\AFF{Key Lab of CNII, MOE, Southeast University, Nanjing, China, \EMAIL{xyf@seu.edu.cn}}
} 

\ABSTRACT{%

We consider the allocation of limited resources to heterogeneous customers who arrive in an online fashion. We would like to allocate the resources ``fairly'', so that no group of customers is marginalized in terms of their overall service rate. We study whether this is possible to do so in an online fashion, and if so, what a good online allocation policy is.

We model this problem using online bipartite matching under stationary arrivals, a fundamental model in the literature typically studied under the objective of maximizing the total number of customers served. We instead study the objective of \textit{maximizing the minimum service} rate across all groups, and propose two notions of fairness: long-run and short-run.

For these fairness objectives, we analyze how competitive online algorithms can be, in comparison to offline algorithms which know the sequence of demands in advance. For long-run fairness, we propose two online heuristics (Sampling and Pooling) which establish asymptotic optimality in different regimes (no specialized supplies, no rare demand types, or imbalanced supply/demand). By contrast, outside \textit{all} of these regimes, we show that the competitive ratio of online algorithms is between 0.632 and 0.732. For short-run fairness, we show for complete bipartite graphs that the competitive ratio of online algorithms is between 0.863 and 0.942; we also derive a probabilistic rejection algorithm which is asymptotically optimal in the total demand.

Depending on the overall scarcity of resources, either our Sampling or Pooling heuristics could be desirable. The most difficult situation for online allocation occurs when the total supply is just enough to serve the total demand, in which case an organization could try to make allocations offline instead.

We simulate our algorithms on a public ride-hailing dataset, which both demonstrates the efficacy of our heuristics and validates our managerial insights.

}%



\maketitle

%

\section{Introduction}

In the online bipartite matching problem,
nodes on one side of a bipartite graph are given in advance,
while nodes on the other side arrive one-by-one.
We refer to the two sets of nodes as \textit{offline} and \textit{online} agents, respectively.
The edges incident to an online agent, which indicate the offline agents eligible to serve it, are revealed upon its arrival.
An online matching algorithm must immediately serve each arriving agent using up to one eligible and unmatched offline agent;
matches once made cannot be rearranged.
The performance of an algorithm is determined by the total
number of matches made,
taking expectations as necessary if there is randomness in the arrivals or the algorithm.
The \textit{competitive ratio} (CR) measures the separation between the performance of online algorithms vs.\ that of a clairvoyant algorithm which knows all of the arrivals in advance.

In this paper, we study online matching problems where performance is instead determined by the \textit{fairness} in service provided to different groups of online agents.
We assume that each online agent belongs to some protected groups, e.g.\ based on race or gender identity, which are observed upon arrival.
To ensure that every group is adequately served, 
we evaluate performance by the \textit{minimum} fraction of demand served over all the groups, defined in two different ways:
\begin{align}
\text{Long-Run Fairness} &=\min_{\text{groups }G}\frac{\E[\text{\# of agents in group $G$ served}]}{\E[\text{\# of arrivals in group $G$}]}; \label{eqn:longRun} \\
\text{Short-Run Fairness} &=\E\left[\min_{\text{groups }G
}
\frac{\E[\text{\# of agents in group $G$ served}]}{\text{\# of arrivals in group $G$}}\right]. \label{eqn:shortRun}
\end{align}

\xhdr{Motivation for Long-Run Fairness.}
The online matching time horizon represents a single day, and the algorithm is audited for fairness after a large number of days $T$ have passed.
In this case, the total number of group-$j$ agents served over all the days will be statistically close to $T$ times the numerator in~\eqref{eqn:longRun},
while the total number of group-$j$ agents to arrive over all the days will be statistically close to $T$ times the denominator.
The audited performance is the minimum of this fraction over all groups $j$.

\xhdr{Motivation for Short-Run Fairness.}
The algorithm is audited for fairness based on the realized arrivals every single day. To avoid impossibility results\footnote{Observe that any deterministic algorithm will yield a fairness of zero during peak hours when there are lots of groups each with a small arrival rate but the total rate is far larger than the serving capacity of offline agents.}, evaluation in the numerator of~\eqref{eqn:shortRun} is based on the \textit{expected} service over any randomness in the algorithm. Interpreted another way, when evaluating Short-Run Fairness, we are allowing for \textit{fractional} allocations to be made on a given day. The overall performance~\eqref{eqn:shortRun} then takes the expectation of the daily audit scores over a large number of days.

Note that our objectives of Long-Run and Short-Run Fairness are percentages between 0 and 1. A guarantee on these percentages does not directly imply that all protected groups will enjoy an equitable level of service; however, these objectives naturally encourage algorithms to allocate the offline agents evenly across the online groups.

We acknowledge that our objectives for fairness at the group level do not address equity at the individual level \citep[see][]{dwork2012fairness,binns2020apparent}; we make no considerations for the most ``deserving'' or ``in need'' agents within each group being served.  Moreover, we are assuming that agents can be correctly labeled and there is no strategic behavior from individuals to obfuscate their groups. Nonetheless, we believe our objectives to be reasonable for large-scale online platforms, on which it has been found that under the current algorithms, agents in certain protected groups are significantly less likely to be served \citep{edelman2017racial,mejia2020transparency}.

We proceed with definitions~\eqref{eqn:longRun}--\eqref{eqn:shortRun} and answer the following questions: 
\begin{enumerate}
\item What is the fairness lost by imposing \textit{non-rejection}, \ie that an online agent must be served (regardless of group) as long as there is an adjacent offline agent with remaining service capacity?
\item In terms of maximizing fairness objectives~\eqref{eqn:longRun} or~\eqref{eqn:shortRun}, computing an optimal online policy may be hard, but can we derive simple, near-optimal online allocation heuristics?
\item What is the competitive ratio, \ie the gap between the objective values~\eqref{eqn:longRun} or~\eqref{eqn:shortRun} achievable by an online algorithm, vs.\ a clairvoyant offline algorithm which knows the arrival sequence in advance?
\end{enumerate}
We believe Questions~1 and~3 to be particularly relevant for online platforms, addressing the design decision of whether incoming agents should be served whenever possible, and how much fairness the platform is losing by serving agents in an online instead of offline fashion.
In this paper, we identify parameter regimes where a simple online heuristic achieves a competitive ratio approaching 100\%, thereby also answering Question~2 in that it is a near-optimal online policy in these regimes.

\subsection{Main Contributions}

In this paper, we assume that online agents arrive following independent Poisson processes with \textit{known}, \textit{homogeneous} rates.
We see the assumption of rates being known as a modeling choice which puts us in the setting of online \textit{stochastic} matching.
On the other hand, our homogeneity assumption, that arrival rates do not change over time, does play a significant role in our results.
We justify this assumption in two ways.
First, note that by rescaling time windows accordingly, one can assume without losing generality that the \textit{total} arrival rate is homogeneous over time.
Therefore, the assumption is only on the \textit{relative} frequencies being unchanging, i.e.\ no group has a tendency to arrive later than other groups, which makes it a much milder assumption.
Second, although this assumption does simplify the problem, i.e.\ by eliminating the need to reserve offline agents for groups with tendencies to arrive later, we see that it still leaves many non-trivial tradeoffs in the design of online algorithms.
Prioritizing model parsimony, we decide to leave non-homogeneous arrival rates outside the scope of this work. 

We now describe our results.
For Long-Run fairness, we show that the competitive ratio of general online algorithms is between $1-1/\sfe\approx0.632$ (\textbf{Theorems~\ref{thm:main-fa-ext1},~\ref{thm:generalGroups}}) and $\sqrt{3}-1\approx0.732$ (\textbf{Theorem~\ref{thm:main-fa-ub}}), while the competitive ratio of non-rejecting online algorithms is exactly 1/2 (\textbf{Theorem~\ref{thm:nonRejUB}}).
Next, we establish that under specific parameter regimes, certain online heuristics achieve a competitive ratio approaching 1:
\begin{enumerate}
\item When there are many copies of every offline agent, an online algorithm which \textit{independently samples} an offline LP solution for each online agent achieves a competitive ratio approaching 1 (\textbf{Theorems~\ref{thm:main-fa-ext1},~\ref{thm:generalGroups}});
\item When all online agent types have a high arrival rate, an online algorithm which \textit{pools and reserves} a set of offline agents to serve each online agent type achieves a competitive ratio approaching 1 (\textbf{Theorem~\ref{thm:main-fa-ext2}});
\item When a demand saturation parameter $s^*$ approaches 0 or $\infty$, the LP sampling algorithm achieves a competitive ratio approaching 1, assuming that every protected group $G$ is \textit{homogeneous}, \ie consists of a single online agent type (\textbf{Theorem~\ref{thm:main-fa-ext1}}).
\end{enumerate}

For Short-Run Fairness, we assume there to be $b$ copies of a \textit{single} offline agent, which can be interpreted as one divisible resource.  We show that the non-rejecting First-Come-First-Serve algorithm achieves a competitive ratio of 0.863 (\textbf{Theorem~\ref{thm:fb-small}}), when the total arrival rate $\Lambda$ of online types is at most 1.
On the other hand, we derive a \textit{probabilistic rejection} algorithm which is asymptotically optimal (\textbf{Theorem~\ref{thm:fb-large}}) as $\Lambda\to\infty$, with $b$ allowed to depend arbitrarily on $\Lambda$.
We note that this algorithm performs rejections using randomness that is \textit{dependent} across agents, making it different from the independent sampling algorithm mentioned earlier.
Finally, we show that the competitive ratio of online algorithms is upper-bounded by 0.942 (\textbf{Theorem~\ref{thm:hard-fb}}), even when $b=1$.

\subsection{Insights about Fairness from our Model and Contributions}

We see our main modeling novelties as: \textit{maximizing the minimum service} ratio provided across different protected groups, when the constituents of each group arrive \textit{online}, and there are \textit{graph} constraints on which supply types can serve which constituents.
We summarize the main takeaways from our model/results which we believe could be useful for increasing fairness in online allocation applications:
\begin{enumerate}
\item When all supply types are common with many units initially available, an online algorithm which rations them \textit{on-the-fly} following an offline allocation can achieve optimal fairness.
\item By contrast, when all protected groups are common with many constituents each, an online algorithm which \textit{pools and pre-reserves} a set of supply units for each group can achieve optimal fairness.
\item For both of the above algorithms, it is important that the online algorithm is permitted to \textit{reject} service to over-served groups.
This shows that in resource allocation, group-level fairness comes at the expense of the first-come-first-serve principle which appears fair to individuals.
\item When the overall demand saturation is very low or very high, it is easy for an online algorithm to achieve the same fairness as an offline algorithm, even if (in the case where demand saturation is very high) this means that all groups are poorly served.
\item In light of the points above, the most difficult situation for achieving fairness online is when: (1) there are ``specialized'' supply units with low availability; (2) there are ``rare'' groups which can only be served by certain specialized supplies; and (3) the total system supply approximately equals the total system demand.
In these situations, since the fairness achieved by an online algorithm can be as bad as $(\sqrt{3}-1)\approx73\%$, we would recommend redesigning the system so that supplies can be allocated \textit{offline} after all demand has arrived.
\item Fairness in expectation (\fa) is much easier to achieve than having to be fair on a given realization (\fb).
\end{enumerate}

\subsection{Experimental Results on Ride-hailing Dataset}
Using a ride-hailing dataset collected from the city of Chicago\footnote{https://data.cityofchicago.org/Transportation/Transportation-Network-Providers-Trips/m6dm-c72p},
we test our heuristics against existing algorithms in the Online Bipartite Matching literature, in some cases adapting them for our Long-Run Fairness objective.
We consider both the general case where protected groups (of riders, based on origin and destination of trip) can consist of heterogeneous types, and the special case where protected groups consist of a single type (\ie a group is defined by a single origin and destination pair).  Our findings are summarized below.
\begin{enumerate}
\item In the case of homogeneous groups, our sampling heuristic always achieves higher Long-Run Fairness than the existing Online Matching algorithms, over a range of choices on how to scale the demand saturation.  Moreover, the general performance of all the algorithms is \textit{exactly consistent} with the managerial insights from our theory---the most difficult situation for achieving fairness in an online fashion arises when the total supply and demand in the system is balanced.  On the other hand, all online algorithms perform better relative to the optimal offline allocation when the supply-demand imbalance increases (in either direction).
\item In the case of heterogeneous groups, online matching using our sampling heuristic is effective if the minimum supply capacity is large, while online matching using our pooling/pre-reserving heuristic is effective if the minimum demand rate is large.  These observations from data are also consistent with our algorithmic guarantees and managerial insights.
\end{enumerate}

\subsection{Organization of Paper}
Theorems~\ref{thm:nonRejUB}--\ref{thm:main-fa-ub} are found in \textbf{Section~\ref{sec:longHomogeneous}}, where we study Long-Run Fairness for homogeneous groups.
Theorems~\ref{thm:generalGroups}--\ref{thm:main-fa-ext2} are found in \textbf{Section~\ref{sec:generalized}}, where we study Long-Run Fairness more generally for heterogeneous groups.
Theorems~\ref{thm:fb-small}--\ref{thm:fb-large} are found in \textbf{Section~\ref{sec:fairb}}, where we study Short-Run Fairness for a single offline type and heterogeneous groups. 
Our experiments can be found in \textbf{Section~\ref{sec:exp}}.

We now discuss some related work in fair operations, which mostly treats fairness as a constraint, instead of an objective to be maximized.
In the concluding \textbf{Section~\ref{sec:conc}}, we discuss some limitations of our approach and the potential side effects of ``maximizing fairness''.

\subsection{Literature Review}

\xhdr{Online Bipartite Matching.}
Online bipartite matching was pioneered by \citet{kvv} and its variants have gained enough interest during the past two decades in the CS community. Based on the arrival setting of online agents, there are three major categories: (1) Adversarial, the arrival sequence is fully is unknown but fixed, see, \eg \cite{buchbinder2007online,mehta2007adwords}; (2) Random arrival order,  the full arrival sequence forms a random permutation over a set of unknown agents, see, \eg \cite{mahdian2011online, karande2011online, goel2008online, devanur2009adwords}; (3) known/unknown distributions,  the stochastic arrivals of online agents follow certain known/unknown distributions.  A special case here is when online arrivals follow  Known Independent and Identical Distributions (KIID), see, \eg \cite{feldman2009online, haeupler2011online, manshadi2012online, jaillet2013online}. Our arrival setting shares the spirit of KIID, though we consider a continuous version instead of discrete.
Recently, \cite{huang2021online} consider the same arrival setting as ours and show that under mild assumptions, the performance of an online algorithm is almost
the same under the two arrival settings (\ie KIID and independent Poisson process).

There is an interesting connection between our model under Long-Run fairness and the \textit{online-side} vertex-weighted online matching under KIID. So far, studies about vertex-weighted online matching all focus on the setting of offline side, \ie all edges incident to any given offline agent share a weight. Examples include
\cite{huang2021online} and
\cite{brubach2016new} under KIID, \cite{huang2018online} under random arrival order, and \cite{aggarwal2011online} under adversarial arrival order.
By contrast, we believe that our analysis and results in Section~\ref{thm:generalGroups} can be applied to the \textit{online-side} vertex-weighted online matching problem, which we leave as future work.


\xhdr{Service Levels in Operations Management.}
Our Long-Run vs.\ Short-Run Fairness objectives distinguish between ``fairness in expectation'' vs.\ ``fairness on every realization'', which correspond to the ``Type-II'' vs.\ ``Type-III'' service rates studied in Operations Management.
A comprehensive discussion of these different ways to measure service (from which our fairness metrics are defined) can be found in the stream of work which studies inventory pooling and supply chain rationing to meet service targets \citep{zhong2018resource,lyu2019capacity,jiang2019achieving}.
However, to our knowledge, this literature has focused on service in an offline setting, with the exception of \citet{li2020personalized}, who incorporate these service definitions into the \textit{constraints} instead of a $\max$-$\min$ objective like we do.

\xhdr{Fair Operations.}
Fairness in operations is a topic of increasing interest and we aim to provide a brief literature review.
Classical works in this area include \citet{bertsimas2011price} and \citet{bertsimas2012efficiency} which define the price of fairness and efficiency-fairness tradeoff, respectively, in an axiomatic fashion.
More recently, ride-sharing platforms have motivated many studies on balancing multiple objectives \citep{lyu2019multi} including fair allocation on the rider side \citep{nanda2019balancing} and income equality on the driver side \citep{xutrade}.
Fair \textit{pricing} to the customer side has been more generally studied in \citet{cohen2019pricing}, while fair allocation in other transportation problems has been studied in \citet{chen2018fairness,chen2020same}.
We note that in the application of \citet{chen2018fairness}, the authors justify prioritizing transportation for certain groups (e.g.\ seniors), instead of balancing fairness across all groups like we do.

More generally, online resource allocation frameworks that can capture fairness have been considered in \cite{balseiro2020regularized,liu2020pond,cheung2020online}.
These papers all derive regret bounds which are sublinear in the number of arrivals, while we derive competitive ratio bounds which hold universally and establish asymptotic optimality in regimes (involving the demand saturation) not previously captured.
However, we should note that our techniques appear to be reliant on the $\max$-$\min$ objective function, while these papers allow for more general functions.



Finally, we should mention while we focus on online bipartite matching, fairness has also been incorporated into other online decision-making questions such as the secretary problem \citep{salem2019closing} or online learning \citep{gupta2019individual,zhang2020fairness}, and other graph-theoretic problems such as influence maximization \citep{tsang2019group} or robust graph covering \citep{rahmattalabi2019exploring}.  In the latter problems, there is a constraint on the fraction of each protected group influenced/covered, which aligns with our proposed $\max$-$\min$ fairness objective.
Very recently, \citet{manshadi2021fair} have studied the online rationing of a single commodity from the perspective of fairness; their model differs from ours in that each customer arrives exactly once, while our objective is maintaining fairness at the aggregate group level.
\section{Model}\label{sec:model}

\xhdr{Graph.} Let $I$ denote the set of offline agents and $J$ denote the set of online types.
For an offline agent $i\in I$, let $\cN_i\subseteq J$ denote the ``neighboring'' online types which $i$ is eligible to serve.
Similarly, for an online type $j\in J$, let $\cN_j\subseteq I$ denote the offline agents eligible to serve $j$.
Each offline agent $i\in I$ has an integer capacity $b_i\ge1$ indicating the maximum number of online agents (with types in $\cN_i$) that $i$ can serve.

\xhdr{Arrivals process.} Agents with each online type $j\in J$ arrive according to an \emph{independent} Poisson process with \emph{homogeneous} rate $\lambda_j>0$, over a time horizon scaled to be $[0,1]$. When an online agent arrives in the time horizon [0,1], its type $j$ is revealed, and an online algorithm must immediately and irrevocably decide whether to serve it using an offline agent $i\in\cN_j$ for which capacity has not been reached.

\xhdr{Protected groups.} There is a set of protected groups $\cG$.
Each group $G\in\cG$ is a subset of $J$, indicating the online agent types that fall under group $G$.
We assume without losing generality that every type $j\in J$ is contained in at least one group (otherwise we could discard and never serve that type); note however that groups can be overlapping.

We refer to the collection of information above (graph, arrival process, protected groups), all of which is known to the algorithm in advance, as an \textit{instance}.

\xhdr{Fairness objectives.} Let $\ALG$ denote a generic online algorithm and allow for algorithms to be randomized. Let $X_j$ be the random variable denoting the number of agents with type $j$ served, for all $j\in J$. For a type $j\in J$, let $A_j$ be the random variable for the number of agents with type $j$ to arrive by the end of the time horizon. For any $\lambda>0$, let $\Pois(\lambda)$ denote a Poisson random variable with mean $\lambda$; note that $A_j$ is then distributionally identical to a $\Pois(\lambda_j)$. Let $\cA$ denote the collection of values $(A_j)_{j\in J}$, which we hereafter call the \textit{arrival vector}. Let $X(G)=\sum_{j\in G}X_j$ and $A(G)=\sum_{j\in G}A_j$ denote
the number of online agents in group $G$ served
and the total number of online agents in group $G$ to arrive,
respectively. Our definitions of \fa and \fb are stated as follows:
\begin{align*}
\fa &=\min_{G\in\cG}\frac{\E_{\cA,\ALG}[X(G)]}{\sum_{j\in G}\lambda_j};&
\fb &=\E_{\cA}\left[\min_{G\in\cG:A(G)>0}\frac{\E_{\ALG}[X(G)|\cA]}{A(G)}\right].
\end{align*}

Here are a few remarks on the above two definitions. (1) Random variables $X_j$ are dependent on both the random arrival vector $\cA$ and any additional random bits used in the algorithm $\ALG$.  In the numerator of \fb, $\E_{\ALG}[X(G)|\cA]$ is a conditional expectation taken over only the randomness in $\ALG$. (2) In \fb, types $j$ with no realized arrivals (for which the denominator $A_j=0$) are ignored. Also, we assume that $\fb=1$ in case all $A_j=0$, \ie no online agents arrive. (3) No inherent relation can be imposed on  \fa and $\fb$. There are  examples  supporting both possibilities that $\fa>\fb$ and $\fa<\fb$; see details in Appendix~\ref{app:fairnessComps}. 

\xhdr{Competitive ratio.}
For any fixed instance (described by $I,J,(\cN_i)_i,(b_i)_i,(\lambda_j)_j$), online algorithm (which may or may not be non-rejecting), and objective (either \fa or \fb), we overload notation and let $\ALG$ denote the objective value of the online algorithm on that instance. Similarly,  we use $\OPT$ to denote an optimal  clairvoyant algorithm and the optimal objective value  when the context is clear. Note that $\OPT$ can set the values of $(X_j)_j$ with advance knowledge of $\cA$.  With a fixed objective in mind, an algorithm $\ALG$ is said to be \textit{$c$-competitive} if $\ALG\ge c\cdot\OPT$ for all possible instances. The maximum possible value over $c\le1$ for which the above holds is called the \textit{competitiveness} of algorithm $\ALG$. The maximum possible competitiveness within a class of online algorithms is called the \textit{competitive ratio} for that class.

\xhdr{Randomized nature of an optimal clairvoyant algorithm}. Consider  the classical (edge-weighted) online bipartite matching  under known IID where the goal is to maximize the total weight of all matches. In that case,  an optimal  clairvoyant algorithm $\OPT$ will aim to optimize the objective on every realized instance and it can always find a deterministic strategy to do so. However,  this may not be true in our problem.  To see this, consider a simple example under \fa where there is one single offline agent with $b=1$ and two online types with $\lam_1=\ep$ and $\lam_2=1$ each of which constitutes its own group. For any realized arrival vector $\cA=(A_1, A_2)$ with $A_1 \ge 1$ and $A_2 \ge 1$, one can show that the strategy of \off on $\cA$ can be characterized as follows: serve $j=1$ and $j=2$ with respective probabilities $p$ and $1-p$, where $p \ge (\sfe-2)/(\sfe-1)\sim 0.418$.  This suggests that $\off$ will have to resort to a randomized strategy on $\cA$---it does not suffice to simply maximizing the objective of $\min \big( \E[X_1]/\ep, \E[X_2]/1\big)$ on every realization of $\cA$.

\xhdr{Some special cases we consider.} In light of the nuanced fairness objectives, along with the randomized nature of the optimal clairvoyant algorithm under $\fa$, analyzing the competitive ratio in our online matching problem is generally challenging.
Moreover, there is no natural technique for bounding the optimal clairvoyant algorithm under $\fb$.
Consequently, there are assumptions which we make in some of our results:
\begin{enumerate}
\item \textit{Homogeneous Groups}: each protected group $G$ consists of a singleton online type $j\in J$, which can \WLOG be assumed to be different for each $G\in\cG$ (otherwise we can eliminate some groups).  In this special case, we refer to groups and types interchangeably. 

\item \textit{A Single Offline Agent}: $I$ consists of a singleton offline agent, which is \WLOG assumed to neighbor every online type.  Note that this single offline agent can still have capacity $b>1$.
\end{enumerate}
We believe Assumption~1 to be mild, in that agents within the same protected group are often homogeneous from the perspective of the online platform (and hence have the same ``type'') anyway.
On the other hand, Assumption~2 restricts us from having different types of offline agents which are eligible to serve different types of online agents, but nonetheless still leaves us with the parsimonious and well-motivated problem of rationing a single resource.

\section{Long-run Fairness with Homogeneous Groups} \label{sec:longHomogeneous}

We first consider \fa under the assumption that each protected group consists of a single type in the matching graph.  Accordingly, in this section we treat types and groups interchangeably. Under this assumption of homogeneous groups, the formulas for long-run and short-run fairness can be simplified as follows:
\begin{align*}
\fa &=\min_{j\in J}\frac{\E_{\cA,\ALG}[X_j]}{\E_{\cA}[A_j]}=\min_{j\in J}\frac{\E_{\cA,\ALG}[X_j]}{\lambda_j}; \\
\fb &=\E_{\cA}\left[\min_{j\in J:A_j>0}\frac{\E_{\ALG}[X_j|\cA]}{A_j}\right].
\end{align*}

First as a warm-up, we see that for \fa, under the further assumption of a single offline agent, the optimal online algorithm is First-Come-First-Serve (\fcfs). It matches all incoming agents to the offline agent as long as capacity is available, and is $1$-competitive. 

\begin{proposition} \label{prop:greedyOptimal}
For  \fa under the two assumptions: (1) homogeneous groups and (2) a single offline agent, \fcfs is a $1$-competitive algorithm.  
\end{proposition}

\begin{myproof}
Suppose that $I$ consists of a single offline agent with capacity $b$.
Let $A$ be the random variable for the total number of online arrivals, in which case FCFS serves the first $\min\{A,b\}$ arrivals.
Conditioned on any value $A>0$, the distribution of online types served is proportional to the arrival rates $\lambda_j$.
That is, for any online type $j\in J$, the expected number of type-$j$ agents served is $\E[\min\{\Pois(\sum_j\lambda_j),b\}]\frac{\lambda_j}{\sum_j\lambda_j}$.
All in all, FCFS achieves a fairness of $\E[\min\{\Pois(\sum_j\lambda_j),b\}]/\sum_j\lambda_j$, which cannot be beaten even by an clairvoyant algorithm since the total number of agents served cannot exceed $\E[\min\{\Pois(\sum_j\lambda_j),b\}]$.
This shows that FCFS is 1-competitive and is also the optimal clairvoyant algorithm.
\end{myproof}

\subsection{Benchmark LP} \label{sec:benchmarkLP}

For instances with multiple heterogeneous offline agents, \fcfs is no longer well-defined, since multiple offline agents could serve an incoming online type.
To guide the choice between offline agents, we write the following LP with variables $x_{ij}$ and $s$.
$x_{ij}$ can be interpreted as the number of times that type $i$ should serve type $j$, while $s$ can be interpreted as the ``scale'' of demand that can be served.
\begin{alignat}{2}
\max & ~~s &&  \label{obj:1} \\
& \sum_{j \in \cN_i} x_{ij} \le b_i  &&~~ \forall i \in I \label{cons:i1} \\ 
&   \sum_{i \in \cN_j} x_{ij} \ge s\cdot \lam_j    && ~~ \forall j \in J \label{cons:g1}\\
& s,x_{ij}\ge0 &&  ~~ \forall (i,j)\in E \label{cons:e1}
\end{alignat}

\begin{lemma} \label{lem:LPbound}
\LP~\eqref{obj:1} is a valid benchmark under \fa, \ie the optimal value of \LP~\eqref{obj:1} offers a valid upper bound for the performance of a clairvoyant algorithm. Therefore, $\off\le\min\{\LP,1\}$.
\end{lemma}

Note that it is important in Lemma~\ref{lem:LPbound} that we also upper bound $\off$ by 1; this will allow us to later establish asymptotic optimality in $s$.

\begin{myproof}
Consider any clairvoyant algorithm.
Let $X_{ij}$ be the random variable for the number of times it uses $i$ to serve $j$, with $X_j=\sum_{i\in\cN_j}X_{ij}$.  Recall that $\OPT=\min_{j \in J}\E[X_j]/\lambda_j$.
It can be checked that setting $x_{ij}=\E[X_{ij}],s=\OPT$ constitutes a feasible LP solution with objective value $\OPT$.
Therefore, $\LP\ge\OPT$, and moreover $1\ge\OPT$ holds by definition, completing the proof.
\end{myproof}

\subsection{Algorithm and Intuition} \label{sec:algIntuition}

In this section, we present an LP-based online sampling algorithm which is $(1-1/\sfe)$-competitive and asymptotically optimal in many parameter regimes. Let $\{x_{ij}^*, s^*\}$ be  an optimal solution to the benchmark LP~\eqref{obj:1}. For all $j\in J$, WLOG assume that $ x_j^* \doteq \sum_{i\in\cN_j}x_{ij}^* = s^*\cdot \lam_j$.\footnote{This is because if $\sum_{i\in\cN_j}x_{ij}^*>s^*\cdot\lambda_j$, then we can re-scale the values of $x^*_{ij}$ by $\frac{s^*\cdot\lambda_j}{\sum_{i\in\cN_j}x_{ij}^*}$, without violating feasibility.} Our LP-based online sampling algorithm, which we dub \nadaps since it depends on the parameter $s^*$, is stated in Algorithm~\ref{alg:nadap}.

\begin{algorithm}[ht!] 
\DontPrintSemicolon
Solve \LP~\eqref{obj:1} to get an optimal solution $\{x_{ij}^*,s^*\}$. \;
Let an online agent (of type) $j$ arrive at time $t$. \;
Sample a neighbor $i \in  \cN_{j}$ with probability $ x^*_{ij}/(s^*\cdot \lam_j)$.  (This is a valid distribution since $\sum_{i \in \cN_j}x_{ij}^*/(s^* \cdot \lam_j)=x_{j}^*/(s^*\cdot \lam_j)=1$.) \label{nalg:step3}\;
If $i$ is safe, \ie $i$ has not reached the capacity, then  assign $i$ to serve $j$; otherwise, reject $j$. \label{nalg:step4}
\caption{An LP-based Sampling algorithm (\nadaps)}
\label{alg:nadap}
\end{algorithm}

\nadaps does not re-sample an offline agent if the first one sampled is unavailable, so it does not share the property of FCFS that an incoming agent is served whenever possible.
The fact that \nadaps sometimes ``rejects'' an incoming agent is important, as illustrated through the following example.

\begin{example}[\textbf{Bad Example}]\label{eg:centralStar}
$J$ consists of
a large number of ``rare types'' $t=1,\ldots,n$ each with $\lambda_t=1/n$ and
a single ``common type'' $0$ with $\lambda_0=n-1$.
$I$ consists of $n$ unit-capacity servers such that each rare type $t=1,\ldots,n$ can only be served by server $t$, but all servers can serve the common type.

It is easy to see that the optimal clairvoyant algorithm gives priority to rare types, and uses servers $t$ for which type $t$ never arrived to serve the common type.  The expected amount of each rare type $t$ served is $1-\sfe^{-1/n}\ge1/n-O(1/n^2)$ while the expected amount of the common type served is at least $n-1-n(1-\sfe^{-1/n})\ge n-2$.  The offline fairness achieved is $1-O(1/n)$.
\end{example}

We first use Example~\ref{eg:centralStar} to show that non-rejecting algorithms cannot be better than 1/2-competitive.

\begin{theorem}\label{thm:nonRejUB}
On Example~\ref{eg:centralStar},
any non-rejecting online algorithm is no more than
$1/2$-competitive relative to the best clairvoyant algorithm.
\end{theorem}

\begin{myproof}
An online algorithm which serves incoming agents whenever possible must have a (randomized) order for available servers to use on the common type.  The rare type which is in position $P \in [n]$ in this order must have an arrival before the $P$'th arrival of the common type $0$, to have any chance of being served.  For a given rare type $t$, let $P_t$ denote the (randomized) position of type $t$ in this order.  For any position $P \in [n]$, let $\arr(P)\in[0,1]$ denote the arrival time of the $P$'th arrival of the common type $0$.  By independence of the Poisson processes for the arrivals of different types, the probability of a rare type $t$ being served is at most
\begin{align} \label{eqn:arrBound}
\E[1-\exp(-\arr(P_t)/n)]
\le\E[\arr(P_t)/n],
\end{align}
which in turn is at most $(\frac{\E[P_t]+1}{n})/n$ for sufficiently large\footnote{This is because as $n\to\infty$, the arrivals of a Poisson process of rate $n$ are evenly spaced in $[0,1]$ w.h.p.} $n$.
Since there must exist a rare type $t$ for which $\E[P_t]\le (n+1)/2$, the fairness of this online algorithm cannot exceed $1/2+O(1/n)$.
\end{myproof}

Now we show that \nadaps overcomes the preceding bound of 1/2 on non-rejecting algorithms, and has a competitiveness which is lower-bounded by $1-1/\sfe$, among other results.  This motivates the need for our online algorithm \nadaps to reject certain types in order to improve fairness for others.

\begin{theorem}\label{thm:main-fa-ext1}
Let $\bmin=\min_{i\in I}b_i$ denote the minimum service capacity of an offline agent $i$.
Under \fa, the competitiveness of online algorithm \nadaps is at least
$$g(b,s^*) \doteq \max\{s^*,1\}\cdot\frac{\E[\min\{\Pois(\bmin/s^*),\bmin\}]}{\bmin},$$ where
\begin{itemize}
\item $g(b,s^*)$ is universally lower bounded at $g(1,1)=1-1/\sfe$;
\item For any fixed $s^*>0$, function $g(b,s^*)$ is minimized at $b=1$, with $\lim_{b\to\infty}g(b,s^*)=1$;
\item For any fixed $b>1$, function $g(b,s^*)$ is minimized at $s^*=1$, with $\lim_{s^*\to0^+}=\lim_{s^*\to\infty}=1$.
\end{itemize}
\end{theorem}

We prove Theorem~\ref{thm:main-fa-ext1} in the next subsection~\ref{sec:thm2}. 
Analysis of when $g(b,s^*)$ is minimized in Theorem~\ref{thm:main-fa-ext1} says that the competitive ratio is worst when the supply types are \textit{specialized} (\ie there are unique offline agents $i$ with $b_i=1$) and the total system supply \textit{exactly meets} total demand (\ie $s^*=1$). Accordingly, our bad Example~\ref{eg:centralStar} satisfied both the criteria that $\min_ib_i=1$ and $s^*=1$.
On the other hand, the asymptotic optimality conditions in Theorem~\ref{thm:main-fa-ext1} say that online algorithms can be \textit{equally fair} as offline algorithms when either:
\begin{itemize}
\item All offline agents have high capacities;
\item Demand is highly saturated (in which case both the online and offline algorithms have poor service);
\item Or demand is very sparse (in which case capacities aren't binding).
\end{itemize}
We believe these conditions for offline algorithms being 1-competitive to be insightful, as summarized in the Introduction.

Finally, we use the same construction in Example~\ref{eg:centralStar} to upper-bound the competitive ratio in general, and to show that our analysis of \nadaps being $(1-1/\sfe)$-competitive is tight.
In both Theorem~\ref{thm:main-fa-ub} and Proposition~\ref{prop:analTight} below, as well as the earlier Theorem~\ref{thm:nonRejUB}, we use the fact that the offline fairness is $1-O(1/n)$ in Example~\ref{eg:centralStar}, and bound from above the online fairness for different algorithms or classes of algorithms.
\begin{theorem}\label{thm:main-fa-ub}
On Example~\ref{eg:centralStar},
any online algorithm is no more than
$(\sqrt{3}-1)$-competitive relative to the best clairvoyant algorithm.
\end{theorem}
\begin{proposition}\label{prop:analTight}
On Example~\ref{eg:centralStar},
\nadaps is no more than
$(1-1/\sfe)$-competitive relative to the best clairvoyant algorithm.
\end{proposition}

Proofs of the above Theorem and Proposition are deferred to Section~\ref{sec:thm3}.

\subsection{Proof of Theorem~\ref{thm:main-fa-ext1}}\label{sec:thm2}

First we use two lemmas to analyze the number of times each online type is served by \nadaps.

\begin{lemma}\label{lem:fa-2}
For each $i \in I$ and $t \in [0,1]$, let $\SF_{it}$ indicate if offline agent $i$ is safe at  the instantaneous point in time $t$ in algorithm \nadaps, \ie $i$ still has remaining capacity at $t$. $\E[\SF_{it}] \ge \Pr[\Pois(b_it/s^*)<b_i]$, for all $i \in I$ and $t \in [0,1]$.
\end{lemma}

\begin{myproof}
Offline agent $i$ is safe at time $t$ if and only if there have been fewer than $b_i$ arrivals before $t$ which sampled $i$.
Such arrivals are Poisson with total rate $\sum_{j \in \cN_i}\lambda_j\cdot\frac{x^*_{ij}}{s^*\cdot\lambda_j}$, which is at most $b_i/s^*$ by LP constraints~\eqref{cons:i1}.
Therefore the number of such arrivals is Poisson with mean at most $b_it/s^*$, completing the proof.
\end{myproof}

\begin{lemma}\label{lem:fa-3}
Let $X^S_j$ be the random number of times $j$ is serviced in \nadaps. Then for all $j\in J$,
\begin{align} \label{eqn:3257}
\frac{\E[X^S_j]}{\lambda_j}\ge s^*\cdot\min_{i\in I}\frac{\E[\min\{\Pois(b_i/s^*),b_i\}]}{b_i}.
\end{align}
\end{lemma}

\begin{myproof}
Consider any $i,j$ for which offline agent $i$ is eligible to serve online type $j$.
Let $X^S_{ij}$ be the random variable for the number of times \nadaps uses $i$ to serve $j$.
$X^S_{ij}$ is incremented whenever: (1) type $j$ arrives (occurring following Poisson process of rate $\lambda_j$); (2) $i$ is sampled (occurring with probability $x^*_{ij}/(s^*\cdot\lambda_j)$); and (3) $i$ is safe (occurring with probability at least $\Pr[\Pois(b_it/s^*)<b_i]$, by Lemma~\ref{lem:fa-2}).
Since these events are mutually independent, we have
\begin{align*}
\E[X^S_{ij}] &\ge\int_0^1\lambda_j\cdot\frac{x^*_{ij}}{s^*\cdot\lambda_j}\cdot\Pr[\Pois(b_it/s^*)<b_i]dt \\
&=\frac{x^*_{ij}}{b_i}\int_0^1\frac{b_i}{s^*}\cdot\Pr[\Pois(b_it/s^*)<b_i]dt =\frac{x^*_{ij}}{b_i}\cdot\E[\min\{\Pois(b_i/s^*),b_i\}].
\end{align*}
The final equality holds because the integral ``counts'' an arrival from a Poisson process of rate $b_i/s^*$ whenever the number of arrivals thus far is less than $b_i$; this equals, in expectation, the number of arrivals from such a process truncated by $b_i$.

Now, for any online type $j\in J$, let $X^S_j=\sum_{i\in\cN_j}X^S_{ij}$ be the random variable for the number of times $\nadaps$ serves $j$.  The previous derivation for $X^S_{ij}$ implies that
\begin{align*}
\E[X^S_j]
\ge\sum_{i\in\cN_j}x^*_{ij}\cdot\frac{\E[\min\{\Pois(b_i/s^*),b_i\}]}{b_i} 
\ge s^*\cdot\lambda_j\cdot\min_{i\in\cN_j}\frac{\E[\min\{\Pois(b_i/s^*),b_i\}]}{b_i}
\end{align*}
where the second inequality uses LP constraint~\eqref{cons:g1}.  This completes the proof.
\end{myproof}

Having derived the expression on the RHS of~\eqref{eqn:3257}, we aim to bound it in terms of simpler expressions of $b_i$ and $s^*$.
Recall that we defined $g(b,s) \doteq \max\{s,1\}\cdot\frac{\E[\min\{\Pois(\bmin/s),\bmin\}]}{\bmin}$ for integer $b \ge 1$ and $s>0$. For any $\lam>0$ and $s>0$, define $h(\lam, s) =\frac{\E[\min\Pois(\lam), \lam s]}{\lam \cdot \min(s,1)}$, a related function we will later use in our analysis. We can verify that $g(b,s)=h(b/s, s)$ and $h(\lam,s)=g(\lam s, s)$. Here are a few properties of $g(b,s)$. 

\begin{lemma}\label{lem:fa-4}
(1) For any fixed $s>0$, $g(b,s)$ is increasing in $b$; (2) For any fixed integer $b \ge 1$, $g(b,s)$ is minimized at $s=1$; (3) For all integers $b \ge 1$ and $s>0$, $g(b,s)\ge g(1,1)=1-1/\sfe$; (4) When $s>1$, $g(b,s) \ge 1-\exp(-b \ln s(1-o(1)) )$, where $o(1)$ vanishes when $s \rightarrow \infty$; (5) When $s=1$, $g(b,1) \ge 1-\frac{1}{\sqrt{2 \pi (b-1)}}$ with $b>1$; (6) When $0<s<1$, $g(b,s) \ge 1-\exp(-\frac{b}{2s}(1-s)^2)$.
\end{lemma}

\begin{myproof} 
Part (1) follows from the simple fact (see \eg \citep{ma2020dynamic}) that $\frac{\E[\min\{\Pois(b/s),b\}]}{b}$ is increasing in $b$.
Part (2) is also easy to see: if $s\le1$, then $g(b,s)=\frac{\E[\min\{\Pois(b/s),b\}]}{b}$ which is decreasing in $s$;
if $s\ge1$, then $g(b,s)=s\frac{\E[\min\{\Pois(b/s),b\}]}{b}$ which is increasing in $s$.
Furthermore, we can derive that
\begin{align*}
\frac{\E[\min\{\Pois(b/s),b\}]}{b}
&=1-\frac{1}{b}\E[\max\{b-\Pois(b/s),0\}] =1-\sum_{k=0}^{b-1}\sfe^{-b/s}\frac{b^{k-1}}{s^kk!}(b-k);
\end{align*}
if $s=1$ then this equals
\begin{align*}
\frac{\E[\min\{\Pois(b/s),b\}]}{b}
&=1-\sum_{k=0}^{b-1}\sfe^{-b}\frac{b^k}{k!}+\sum_{k=1}^{b-1}\sfe^{-b}\frac{b^{k-1}}{(k-1)!} =1-\sfe^{-b}\frac{b^{b-1}}{(b-1)!} =g(b,1).
\end{align*}
It can be verified that $g(b,1) $ gets minimized at $b=1$ with $g(1,1)=1-1/\sfe$.  For $b>1$,
$$
g(b,1) \ge 1-\sfe^{-b}\frac{b^{b-1}}{\frac{(b-1)^{b-1}}{\sfe^{b-1}}\sqrt{2\pi(b-1)}}
= 1-\frac{1}{\sfe}(1+\frac{1}{b-1})^{b-1}\frac{1}{\sqrt{2\pi(b-1)}}
\ge 1-\frac{1}{\sqrt{2\pi(b-1)}}
$$
where we have used Stirling's approximation in the first inequality. This establishes Part (3) and Part (5).

Now we show Parts (4) and (6). Recall that $h(\lam, s) =\frac{\E[\min\Pois(\lam), \lam s]}{\lam \cdot \min(s,1)}$ and $g(b,s)=h(b/s, s)$. Consider the first case when $s>1$. We see that 
\begin{align*}
h(\lam,s)&= \frac{\E[\min(\Pois(\lam), \lam s)]}{\lam} \ge
\frac{1}{\lam} \sum_{k=1}^{\lam s} \frac{\sfe^{-\lam} \lam^k k }{ k! } =  \sum_{k=0}^{\lam s-1} \frac{\sfe^{-\lam} \lam^k  }{ k! }=1-\Pr[\Pois(\lam) \ge \lam s ]\\
& \ge 1-\exp \Big( -\lam \frac{\ln s \cdot (s-1)^2}{s} (1-o(1))\Big).
\end{align*}
The last inequality is due to the upper tail bound of a Poisson random variable as shown by~\cite{pois-tail}, where $o(1)=\Theta(1/\ln s)$ is a vanishing term when $s$ is large. Thus, since $g(b,s)=h(b/s,s)$, we see $g(b,s) \ge 1-\exp(-b \cdot \ln s \cdot (1-1/s)^2 (1-o(1)))$, completing Part (4).

Similarly for $s<1$, we have 
\begin{align*}
h(\lam,s)&= \frac{\E[\min(\Pois(\lam), \lam s)]}{\lam s} \ge
\frac{\lam s}{\lam s} \sum_{k=\lam s}^{\infty} \frac{\sfe^{-\lam} \lam^k  }{ k! } =1-\Pr[\Pois(\lam) < \lam s ] \ge 1-\exp\Big(-\frac{\lam (1-s)^2}{2}\Big).
\end{align*}
The last inequality is due to~\citep{pois-tail}. Thus, by replacing $\lam$ with $b/s$, we establish Part (6). 
\end{myproof}

\begin{myproof}[Proof of Theorem~\ref{thm:main-fa-ext1}]
By Lemma~\ref{lem:LPbound}, $\OPT\le\min\{s^*,1\}$.
By Lemma~\ref{lem:fa-3}, the fairness of \nadaps under \fa is at least $\frac{\E[X^S_j]}{\lambda_j}\ge s^*\cdot\min_{i\in I}\frac{\E[\min\{\Pois(b_i/s^*),b_i\}]}{b_i}$.
By Lemma~\ref{lem:fa-4} Part~(1), this is lower-bounded by $s^*\cdot\frac{\E[\min\{\Pois(\bmin/s^*),\bmin\}]}{\bmin}$, with $\bmin=\min_i b_i$. Putting these statements together, we see that the competitive ratio is lower-bounded by
\[
\frac{s^*}{\min\{s^*,1\}}\cdot\frac{\E[\min\{\Pois(\bmin/s^*),\bmin\}]}{\bmin}
=g(b,s^*)\]

All of the properties about $g(b,s^*)$  follow directly from Lemma~\ref{lem:fa-4}, 
with the asymptotic behavior when $b\to\infty$, $s^*\to0^+$, or $s^*\to\infty$ following from the bounds given in parts~(4)--(6) of Lemma~\ref{lem:fa-4}.
\end{myproof}

\subsection{Proofs of Theorem~\ref{thm:main-fa-ub} and Proposition~\ref{prop:analTight}}\label{sec:thm3}
\begin{myproof}[Proof of Theorem~\ref{thm:main-fa-ub}.]
On Example~\ref{eg:centralStar}, any online algorithm which is going to reject the common type is better off doing so sooner rather than later, since an earlier rejection allows more time to observe which rare types arrive, and give those types priority.
For any $\tau\in[0,1]$, suppose that the online algorithm, denoted by $\ALG(\tau)$, starts accepting common types after time $\tau$.

The online algorithm must have some (possibly randomized) order of offline servers to use when it wants to serve the common type.
The rare type whose corresponding offline server is in position $P\in[n]$ in this order must have an arrival before the $P$'th arrival of the common type \textit{after time $\tau$}, to have any hope of being served.
Counting from time $\tau$, the $P$'th arrival of the common type will occur before $\tau+\frac{P+1}{n}$ w.h.p.\ as $n\to\infty$.  As a result, the probability of this rare type being served is at most
$$
1-\exp(-\frac{\min\{\tau+\frac{P+1}{n},1\}}{n})\le\frac{\min\{\tau+\frac{P+1}{n},1\}}{n}.
$$
As $n\to\infty$, the average value of the RHS expression over $P=1,\ldots,n$ is
\begin{align*}
\frac{1}{n}\int_0^1\min\{\tau+z,1\}dz=\frac{1}{n}(\tau+\frac{1}{2}-\frac{1}{2}\tau^2).
\end{align*}
Therefore, even using a randomized order, there must exist a rare type whose probability of being served is at most $\frac{1}{n}(\tau+\frac{1}{2}-\frac{1}{2}\tau^2)$.
Meanwhile, for any $\tau$, the expected number of common types served can be at most $(n-1)(1-\tau)$.
Since the arrival rates for rare and common types are $\frac{1}{n}$ and $n-1$ respectively, the fairness of the online algorithm cannot exceed $\min\{\tau+\frac{1}{2}-\frac{1}{2}\tau^2,1-\tau\}$.

We can verify that the fairness of the online algorithm is maximized at $\tau=2-\sqrt{3}$, in which case it equals $\sqrt{3}-1$.
Meanwhile, for Example~\ref{eg:centralStar}, an clairvoyant algorithm can achieve a fairness of 1.
This completes the proof.
\end{myproof}

\begin{myproof}[Proof of Proposition~\ref{prop:analTight}.]
The optimal LP solution sets $x^*_{t,t}=1/n$ and $x^*_{t,0}=1-1/n$ for each $t=1,\ldots,n$, with $s^*=1$.
As a result, an offline agent $t \in [n]$ hence faces a demand which is $\Pois(1)$.
Offline agent $t$ successfully serves a demand with probability $1-1/\sfe$, and conditioned on this, the probability of that demand being of rare type $t$ (instead of the common type $0$) is $1/n$.
Thus, for any rare type $t \in [n]$, we have $\E[X_t]/\lam_t=1-1/\sfe$, where $X_t$ denotes the random number of times type $t$ is serviced. Therefore, and under \fa, \nadaps achieves a fairness of at most $1-1/\sfe$.
Meanwhile, on Example~\ref{eg:centralStar}, it is possible for an clairvoyant algorithm to achieve a \fa of $1-O(1/n)$, completing the proof.
\end{myproof}

\section{Long-run Fairness with Heterogeneous Groups} \label{sec:generalized}

In this section we consider the general model described in Section~\ref{sec:model} where protected groups can consist of multiple different types and potentially overlap with each other.
In Subsection~\ref{sec:reserve}, we introduce another online algorithm \reserve based on inventory pooling which is asymptotically optimal if all online types are common, something not achieved by the previous algorithm \nadap.
All missing proofs from this section are deferred to Appendix~\ref{app:pfsGeneralized}.

Here is the updated version of Benchmark LP for \fa without the assumption of homogeneous groups.
\begin{alignat}{2}
\max & ~~s && \label{obj:general}  \\ 
& \sum_{j \in \cN_i} x_{ij} \le b_i  &&~~ \forall i \in I \nonumber \\ 
\sum_{j\in G} &   \sum_{i \in \cN_j} x_{ij} \ge s\cdot \sum_{j\in G}\lam_j    && ~~ \forall G \in \cG \nonumber \\ 
& \sum_{i\in\cN_j}x_{ij}\le\lambda_j && ~~ \forall j\in J \label{cons:added}\\
& s,x_{ij}\ge0 &&  ~~ \forall (i,j)\in E \nonumber
\end{alignat}
Note that we have added a new set of constraints~\eqref{cons:added}, which are clearly valid for any clairvoyant algorithm since the constraints hold on every sample path based on the realized number of arrivals and services.  Therefore, if we let $\{x_{ij}^*,s^*\}$ denote an optimal solution to the LP, then $\OPT\le s^*$. We now state our generalization of algorithm \nadaps, which we dub \nadap in light of fact that its performance no longer depends on the optimal $s$ value.

\begin{algorithm}[ht!] 
\DontPrintSemicolon
Solve \LP~\eqref{obj:1} to get an optimal solution $\{x_{ij}^*\}$. \;
Let an online agent (of type) $j$ arrive at time $t$. \;
Sample a neighbor $i \in  \cN_{j}$ with probability $ x^*_{ij}/\lam_j$.  (This is a valid distribution due to Constraint~\eqref{cons:added}.) \label{nalg2:step3}\;
If $i$ is safe, \ie $i$ has not reached the capacity, then  assign $i$ to serve $j$; otherwise, reject $j$. \label{nalg2:step4}
\caption{Generalized LP-based Sampling algorithm (\nadap)}
\label{alg2}
\end{algorithm}
Note that $\nadap$ will reject an online agent immediately with probability $1-\sum_{i\in\cN_j}x^*_{ij}/\lambda_j$, and will also reject it if the first sampled offline agent has reached capacity.

Before stating our generalized theorem for the performance of \nadap, a few remarks on why the competitiveness no longer depends on $s^*$ should be made.
Recall that in Theorem~\ref{thm:main-fa-ext1}, $s^*$ was interpreted as the ``scale'' of demand which can be served, and the competitiveness approached 1 if $s^*\to\infty$ or $s^*\to0^+$.
However, in the generalized model with groups, $s^*$ no longer has this interpretation and these statements about asymptotic optimality no longer hold.
We provide examples below.
\begin{itemize}
\item First, $s^*\to\infty$ is no longer possible, because $s\le1$ is implied by constraints~\eqref{cons:added}.
On the other hand, if we do not add these constraints, then the LP has an unbounded gap, as demonstrated by the following example.
There is a single group consisting of $n$ types with arrival rates $1$.
One type is connected to an offline agent with capacity $n$; the other types are connected to no offline agents.
Without constraints~\eqref{cons:added}, the LP would be able to ``overserve'' the first type and achieve a fairness of 1; any actual algorithm would have a fairness at most $1/n$.
All in all, in the generalized model, it is no longer possible to allow an $s$ which is greater than 1.
\item If $s^*\to0^+$, it is no longer the case that online algorithms can achieve a fairness of $s^*$, as demonstrated by the following example.
There is a single group consisting of 2 types; one with arrival rate 1 and the other with arrival rate $\lambda\to\infty$.
Each type is connected its own offline agent with capacity 1.
In this case $s^*=2/(1+\lambda)$, which approaches 0.
However, an online algorithm makes in expectation only $1+(1-1/\sfe)$ services, achieving fairness $(2-1/\sfe)/(1+\lambda)$.
\end{itemize}

\begin{theorem} \label{thm:generalGroups}
The competitiveness of \nadap is at least $1-\sfe^{-b}\frac{b^b}{b!}$, which is increasing in $b$ (recall that $b=\min_{i\in I}b_i$) and approaches 1 as $b\to\infty$.
\end{theorem}

\subsection{\fa when All Online Types are Common} \label{sec:reserve}
In this section we introduce another regime in which online algorithms are 1-competitive---the regime where all online types are common, \ie have high arrival rates.
However, this regime requires a different algorithm, which we now motivate using the following example.

\begin{example}\label{eg:poolSupply}
$J$ consists of a single type $1$ with $\lambda_1=n$ and $I$ consists of $n$ separate servers each with unit capacity.
Using \nadap, each server faces a separate demand according to a Poisson process of rate 1, and successfully serves demand with probability $1-1/\sfe$.  The total expected demand served is $n(1-1/\sfe)$.
However, an algorithm which adaptively chooses an available server and never rejects incoming demand as long as a server is available serves a total expected demand of $\E[\min\{\Pois(n),n\}]$.
As $n\to\infty$, the \fa of the adaptive algorithm approaches 1, while the \fa of \nadap is stuck at $1-1/\sfe$.
\end{example}

\nadap did not improve on this example even when the arrival rate approached $\infty$ because it did not ``pool'' the servers in order to reduce the variance in demand served.
Motivated by this example, we now introduce an algorithm \reserve which pre-assigns the capacity that will be used to serve each online type.
In general, offline agents could be adjacent to many online types and may not be as straight-forward to assign as in Example~\ref{eg:poolSupply}; however we make use of the same generalized LP from Section~\ref{sec:generalized} along with the dependent rounding procedure of \citet{gandhi2006dependent} to generate a randomized assignment.

\begin{algorithm}[ht!]
\DontPrintSemicolon
Split and re-index offline agents as necessary so that $b_i=1$ for all $i\in I$.\;
Solve \LP~\eqref{obj:1} to get an optimal solution $\{x_{ij}^*,s^*\}$, and define $x^*_j=\sum_{i\in\cN_j}x_{ij}^*$ for all $j\in J$. Note that $x^*_j\le\lambda_j$ for all $j$, by constraints~\eqref{cons:added}.\;
Round the LP solution to get binary variables $X^R_{ij}$ such that $\sum_{j\in\cN_i}X^R_{ij}\le1$ for all $i\in I$.\;
For each online type $j$, reserve the offline agents $\{i:X^R_{ij}=1\}$ exclusively for serving $j$, and match them to incoming type-$j$ agents in any first-come-first-serve manner.
\caption{Alternate Algorithm which Pre-reserves Capacities (\reserve)}
\end{algorithm}
For all $j$, let $\serve(j)$ denote the set $\{i:X^R_{ij}=1\}$, which is generally randomized.  By \citet{gandhi2006dependent}, it is possible to do the rounding in Step~3 so that the sets $\{\serve(j):j\in J\}$ are always mutually disjoint, and $|\serve(j)|\in\{\lfloor x^*_j\rfloor,\lceil x^*_j\rceil\}$ for all $j$ with $\E[|\serve(j)|]=x^*_j$.

\begin{theorem} \label{thm:main-fa-ext2}
Let $\lambda=\min_{j\in J}\lambda_j$ denote the minimum arrival rate of an online type $j$.
Under \fa, the competitiveness of online algorithm \reserve is at least
$
1-\sfe^{-\lambda}\frac{\lambda^{\lambda}}{\lambda!},
$
which approaches 1 as $\lambda\to\infty$.
\end{theorem}

Note that the dependence on $b$ in Theorem~\ref{thm:generalGroups} is identical to the dependence on $\lambda$ in Theorem~\ref{thm:main-fa-ext2} even though the algorithms analyzed, \nadap and \reserve, are different.
It would be interesting future work to consider a hybrid between \nadap and \reserve which can pool the supply for common demand types (as in \reserve) while assigning via sampling for rare demand types (as in \nadap).
\section{Short-run Fairness with a Single Offline Type} \label{sec:fairb}

In this section we consider the competitive ratio of online algorithms under Short-run Fairness.
We note that upper-bounding the performance of an offline algorithm appears to be very difficult under \fb (the LP benchmark is no longer an upper bound), which is why we focus on the special case of a \textit{single} offline agent with service capacity $b$.  Even in this special case, the optimal online algorithm which maximizes \fb is complex to characterize (contrast this with Proposition~\ref{prop:greedyOptimal}, which says that \fcfs maximizes \fa with a single offline agent).

Recall that $\cI(b,\lam)$ refers to an instance that has a single offline agent with service capacity $b$, and a total online arrival rate of $\lam$, where the competitive ratios will depend on $\lam$.
Our analysis in this section holds if protected groups can consist of heterogeneous types and overlap with each other.
All missing proofs from this section are deferred to Appendix~\ref{app:fb}.

\subsection{$\cI(b,\lam)$ with $\lam \le 1$ }\label{sec:fb-small}

First we consider instances $\cI(b,\lam)$ with $\lam\le1$.  In this regime, we show that \fcfs is 0.863-competitive.

\begin{theorem} \label{thm:fb-small}
\fcfs is $0.863$-competitive for $\cI(b,\lam)$ with $\lam \le 1$.
\end{theorem}

On the other hand, we show that any online algorithm, even one which can selectively reject online types in a probabilistic and/or adaptive fashion, cannot be more than 0.942-competitive.  We emphasize that establishing this \textit{separation} between online vs.\ offline algorithms is complex because it requires characterizing an optimal online algorithm for \fb.  In Appendix~\ref{app:fb} we prove Theorem~\ref{thm:hard-fb} by analyzing the Bellman equations that govern the optimal online algorithm, and deriving upper bounds which lead to a solvable differential equation.

\begin{theorem} \label{thm:hard-fb}
No algorithm can be more than $0.942$-competitive for $\cI(b,\lam)$ with $b=\lam=1$. 
\end{theorem}

\subsection{$\cI(b,\lam)$ with $\lam \gg1$}\label{sec:fb-large}



We consider $\cI(b,\lam)$ with $\lam \gg 1$ here. Throughout this subsection, we define $\kap \doteq b/\lam$. Here is the result for the performance of an optimal clairvoyant algorithm, denoted by \off, when $\kap<1$.

\begin{lemma}\label{lem:off-fb-ext}
Consider  $\cI(b,\lam)$ with $\kap<1$ under \fb. We have $\off \le \kap \big(1+1/\lam+o(1/\lam)\big)$.
\end{lemma}

Now we present an algorithm which shares the spirit as \reserve as shown in Section~\ref{sec:reserve}. It is mainly based on the technique of dependent rounding (DR) as shown by~\cite{gandhi2006dependent}. We call it the Probabilistic-Rejection algorithm.  Let $K=\lfloor \lam (1+\ep)\rfloor$ where $\ep$ is a parameter to choose later.  The algorithm is stated as follows.

\begin{algorithm}[ht!]
\caption{The Probabilistic-Rejection algorithm for $\cI(b,\lam)$ under \fb.}
\label{alg:fb-3}
\DontPrintSemicolon
Apply dependent rounding to the vector $\x=(b/K) \cdot \bo$, which has $K$ identical entries all equal $b/K$, where $K=\lfloor \lam (1+\ep)\rfloor$. Let $ (Y_k)_k\in \{0,1\}^K$ be the random output vector. \;
Suppose an online agent of type $j$ arrives and suppose it is the $k$'th arrival among all types of online agents.  \;
If $k \le K$ and $Y_k=1$, then serve the incoming type-$j$ agent if possible;  otherwise, reject $j$. 
\end{algorithm}

Theorem~\ref{thm:fb-large} shows that online algorithms can be 1-competitive as $\lambda\to\infty$, even if the service capacity $b$ is increasing at the same time.  Depending on whether $k=b/\lambda$ is greater than 1, the probabilistic rejection probabilities have to be chosen differently.  Also, we note that due to dependent rounding, Algorithm~\ref{alg:fb-3} is less likely to reject an agent if other agents have already been rejected, distributing equal opportunity among the first $K$ arrivals to be served.
This dependent rounding makes it different from Algorithm~\ref{alg:nadap} and similar algorithms in the literature.

\begin{theorem}\label{thm:fb-large}
Consider $\cI(b,\lam)$ under \fb. (1) If $\kap>1$, then by choosing $\ep=\kap-1$, the competitiveness of Algorithm~\ref{alg:fb-3} at least $1-\exp\big(-\lam (\kap-1)^2/(2 \kap)\big)$, which approaches 1 as $\lambda\to\infty$.
(2) If $\kap \le 1$, then by choosing $\ep=\sqrt{\ln \lam /\lam}$, the competitiveness of Algorithm~\ref{alg:fb-3} is at least $1-\sqrt{\ln \lam /\lam}(1+o(1))$, where $o(1)$ is a vanishing term (and hence the competitiveness approaches 1) as $\lam \rightarrow \infty$.
\end{theorem}

\section{Experimental Results on Ride-hailing Dataset}\label{sec:exp}


\xhdr{Preprocessing.}
We test our algorithms \nadaps, \nadap, and \reserve on a ride-hailing dataset\footnote{https://data.cityofchicago.org/Transportation/Transportation-Network-Providers-Trips/m6dm-c72p}, which is reported by the Transportation Network Providers in the city of Chicago.
As of the end of October, $2020$, there were $169$ million trips in total.
Each trip record is made up of $21$ columns, including the unique identifier for the trip, the times when the trip started and ended, the origin (pick-up) and destination (drop-off) locations for the passenger, and the fare
for the trip.
Following~\cite{nanda2019balancing,xutrade}, we focus on the online service rates for riders arriving dynamically, while drivers are assumed to be offline agents. Our goal is to maximize the \emph{(long-run) group fairness} among all riders' groups.
Note that Chicago is made up of $76$ pre-defined community areas which do not overlap with each other, and hence we can categorize all trips according to which of these well-defined areas they start/end in.
In our case, we define a rider-group for each of the $76$ areas and assume each rider belongs to the group identified as her \emph{destination} community area (usually marked as her residence or working area). Recall that our metric of (long-run) group fairness is defined as the minimum fraction of demand served over all groups.


We construct the input bipartite graph as follows.
For each (origin, destination)-pair (of which there are $76^2$ possibilities when categorized by the areas), we create a rider type $j$, and set its arrival rate $\lam_j$ as the average number of records from that origin to that destination over the days of September, 2020 between 18:00 and 19:00.  (We arbitrarily chose this time window over which the traffic conditions tended to be relatively stationary over September, 2020.)
We keep the 484 (origin, destination)-pairs with the highest frequencies.
In the case of homogeneous groups, we define a group for each rider type; in the case of heterogeneous groups, we put all rider types with the same destination area into the same group.  For each origin area, we create a driver type $i$ with service capacity $b_i$ equal to the average number of trip records from that starting area.
For each pair of driver and rider types, we add an edge between them if and only if the driver type's area is the same as the rider type's origin area.

\xhdr{Algorithms.} We compare \nadaps, \nadap, and \reserve against the following existing algorithms. (a) \gre: Always try to assign an arriving online agent to an available neighbor who has the most service capacity; break ties uniformly at random. (b) \rank: Fix a uniform random priority ordering of offline agents in advance; assign each arriving online agent to an available neighbor with the highest priority (if possible). (c) \mgs: the sampling strategy from \citet{manshadi2012online} which generates \textit{two} candidate neighbors upon the arrival of an online agent.  We exclude \mgs in our experiments for homogeneous groups since it requires the constraints $\sum_{i\in\cN_j}x_{ij}^*\le\lambda_j$, which is absent from Benchmark \LP~\eqref{obj:1} for homogeneous groups.  We also note that the state-of-the-art algorithms for online matching under KIID in \citet{brubach2016new,jaillet2013online} cannot be applied on our experimental instances, since they assume unit integral arrival rates (\ie all $\lam_j=1$) for all online types, which does not hold here.

\begin{figure}[t!]
  \centering
  {\includegraphics[width=0.45 \columnwidth]{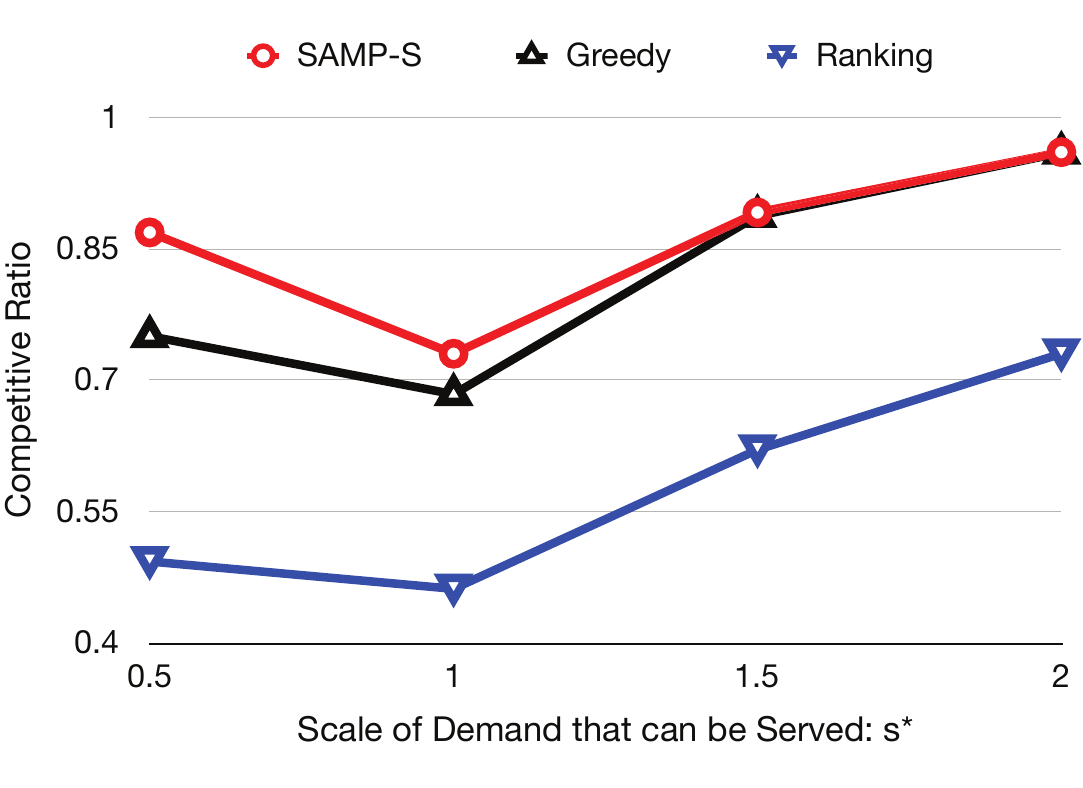}}
  \vspace{-4mm}
     \caption{Comparison of performance of \nadaps, \gre, and \rank under the metric of Long-run Fairness with homogeneous groups. The scale of demand that can be served $s^*$ takes values in $\{0.5,1,1.5,2\}$.}
     \label{fig:homo}
  \vspace{-3mm}
\end{figure}

\begin{figure}[ht!]
  \centering
  {\includegraphics[width=0.45 \columnwidth]{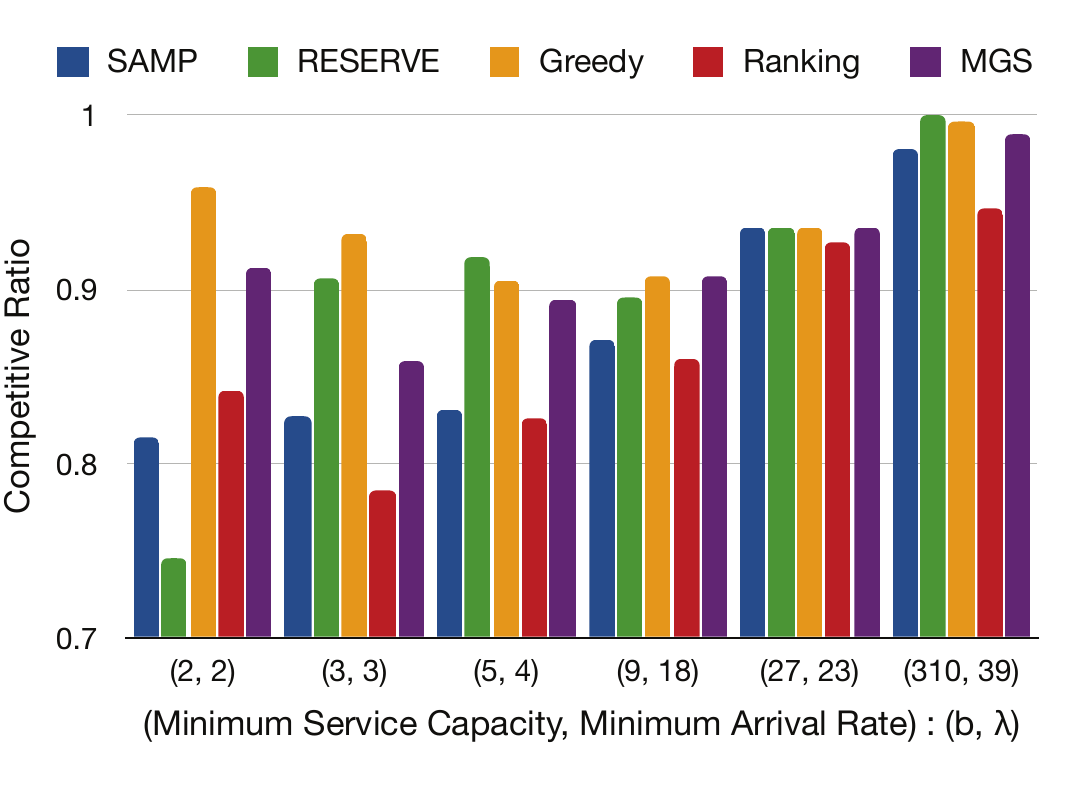}}
      \vspace{-4mm}
     \caption{Comparison of performance  of \nadap and \reserve with \gre, \rank, and \mgs on Long-Run fairness with heterogeneous groups, with $(b,\lam)$ (the minimum service capacity and the minimum arrival rate) taking values in $\{(2,2),(3,3),(5,4),(9,8),(27,23),(310,39)\}$.}
     \label{fig:hete_cmp}
     \vspace{-3mm}
\end{figure}

\xhdr{Results.}
For the case of homogeneous groups, we compare the performance of our \nadaps heuristic against \gre and \rank, allowing the ``scale of serviceable demand'' parameter $s^*$ to take values in $\{0.5,1,1.5,2\}$ by adjusting all offline agents' capacities proportionally. For the case of heterogeneous groups,
we compare the performance of \nadap and \reserve against \gre, \rank, and \mgs, when the two parameters, the minimum service capacity $b$ and the minimum arrival rate $\lambda$, take values from $\{(2,2),(3,3),(5,4),(9,8),(27,23),(310,39)\}$. For all instances, we run $1000$ trials and take the average as the final performance. The competitive ratios are computed by comparing the averaged performance of the algorithm to the optimal value of benchmark \LP~\eqref{obj:1} (homogeneous case) and~\LP~\eqref{obj:general} (heterogeneous case).
The results are shown in \textbf{Figure~\ref{fig:homo}} (homogeneous case) and \textbf{Figure~\ref{fig:hete_cmp}} (heterogeneous case).  We note that since in each plot all algorithms are divided by the same LP objective, competitive ratio is proportional to performance in terms of the objective of Long-Run Fairness.

\xhdr{Discussion.}
Figure~\ref{fig:homo} shows that for the homogeneous case, our \nadaps heuristic outperforms the other two algorithms \gre and \rank universally over all choices of $s^*$.
Also note that the competitive ratio of \nadaps always stays above $1-1/\sfe$, and is worst when $s^*=1$ (when total supply meets total demand), and approaches 1 as $s^*$ increases.
All of these behaviors are \textit{exactly consistent} with our result for \nadaps in Theorem~\ref{thm:main-fa-ext1}, and these behaviors even generalize to the other two baseline algorithms \gre and \rank.
This corroborates using data from practice our managerial insight from Theorem~\ref{thm:main-fa-ext1}---fairness is hardest to maintain in an online fashion when the total supply and demand in the system are roughly balanced; on the other hand, online algorithms can be as fair as the optimal offline allocation as this imbalance grows.

Figure~\ref{fig:hete_cmp} shows that for the heterogeneous case, \gre is the best algorithm when the minimum service capacity ($b$) and the minimum arrival rate ($\lam$) are small.  However, our heuristics \nadap and \reserve quickly catch up and become best-performing once $b$ and $\lam$ exceed 5, with \reserve performing particularly well especially when $b>\lam$.  The algorithms generally have competitive ratios approaching 1 as $b$ and $\lam$ further increase.  This corroborates our managerial insights from Theorems~\ref{thm:generalGroups}--\ref{thm:main-fa-ext2}---online algorithms can be as fair as the optimal offline allocation as either $b$ or $\lam$ grows, assuming the correct algorithm (\nadap in the case of $b\to\infty$, \reserve in the case of $\lambda\to\infty$) is used.

\section{Conclusion and Reservations} \label{sec:conc}

We propose algorithms for maintaining statistical parity in the service rates provided to different groups, when agents arrive sequentially and some groups of agents are more easily serviced than others.
We believe this has the potential to make a positive impact on \eg sharing economy platforms, where our algorithms will give priority to under-served groups when matching agents, thereby boosting their rates of service.
However, we should admit that our algorithms do not address any underlying discrimination issues of why those groups were less commonly served by hosts/drivers in the first place. Also, our algorithms are only ``fair'' with respect to the group-fairness metrics we defined.
Our model does not capture fairness at the individual level, which is more nuanced in our problem, because the agents arrive sequentially.
Relatedly, our algorithms could have the negative consequence of causing ``unfairness'' by violating the first-come-first-serve principle, since sometimes earlier-arriving agents are rejected in order to preserve capacity for later-arriving agents who may belong to protected groups.

\ACKNOWLEDGMENT{%
}

%
%
%


\bibliographystyle{informs2014} 
\bibliography{stable_ref} 


\newpage
\begin{APPENDICES}

\section{Missing Proofs from Section~\ref{sec:model}} \label{app:fairnessComps}

Here are examples showing it is possible that $\fb>\fa$ and $\fb<\fa$.
\begin{example}\label{exam:1}
Consider a simple example where we have one single offline agent and one single online type with $b=\lam=1$. Consider the algorithm \fcfs: serve the online agent whenever it arrives.

Let $A \sim \Pois(1)$ be the number of arrivals of online agents. Observe that $\fa=\E[X]=\Pr[A \ge 1]=1-1/\sfe$. Note that when $A=0$, we have $\fb=1$. Thus, we can verify that
\begin{align*}
\fb&= \Pr[A=0]+\sum_{k=1}^{\infty} \frac{\Pr[A=k]}{k}> \sfe^{-1}+\sum_{k=1}^\infty \frac{\sfe^{-1}}{k!}\frac{1}{k+1}
\\
&=\sfe^{-1}\bp{1+ \sum_{k=2}^\infty \frac{1}{k!}}
=\sfe^{-1}\bp{1+\sfe-2}=1-1/\sfe=\fa.
\end{align*}
\end{example}
Thus we claim that it is possible that $\fb>\fa$.

\begin{example}\label{exam:2}
Consider a simple example where we have one single offline agent and one single online type with $b=1$ and an online arrival rate of $\lam$. Consider such an algorithm featured by a threshold $k$ as follows: serve the online agent only when it arrives for the $k$th time. In other words, ignore it for the first $k-1$ arrivals. Let $A \sim \Pois(\lam)$ denote the number of online arrivals.

Take $\lam=10$ and $k=11$. We can verify that (1) $\fa=\frac{\Pr[A \ge k]}{\lam}$; (2) 
\[
\fb = \Pr[A=0]+\sum_{\ell=k}^{\infty} \Pr[A=\ell]/\ell <\sfe^{-\lam}+\Pr[A\ge k]/k <\Pr[A \ge k]/\lam=\fa.
\]
Thus, we claim that it is possible that $\fb<\fa$.
\end{example}

\section{Missing Proofs from Section~\ref{sec:generalized}} \label{app:pfsGeneralized}

\begin{myproof}[Proof of Theorem~\ref{thm:generalGroups}.]
We provide a terse argument since detailed logic can be found in Lemmas~\ref{lem:fa-2}--\ref{lem:fa-3}.
The incoming demand flow to an offline agent $i\in I$ is Poisson with rate $\sum_{j\in\cN_i}\lambda_j\frac{x^*_{ij}}{\lambda_j}$, which is at most $b_i$ by LP feasibility.
Therefore, the capacity of any offline agent $i$ has not been reached at time $t$ with probability at least
$\Pr[\Pois(b_it)<b_i]$.
Using this fact, the expected number of times offline agent $i$ serves online type $j$ is at least
\begin{align*}
\int_0^1\lambda_j\frac{x^*_{ij}}{\lambda_j}\Pr[\Pois(b_it)<b_i]dt
=&\frac{x^*_{ij}}{b_i}\int_0^1b_i\Pr[\Pois(b_it)<b_i]dt \\
=&x^*_{ij}\frac{\E[\min\{\Pois(b_i),b_i\}]}{b_i}
\end{align*}
Applying Lemma~\ref{lem:fa-4} twice, the expected total number of times a group $G\in\cG$ is served is at least
\begin{align*}
\sum_{j\in G}\sum_{i\in\cN_j}x^*_{ij}\frac{\E[\min\{\Pois(b_i),b_i\}]}{b_i}
\ge&\frac{\E[\min\{\Pois(b),b\}]}{b}\sum_{j\in G}\sum_{i\in\cN_j}x^*_{ij} \\
=&(1-\sfe^{-b}\frac{b^b}{b!})\sum_{j\in G}\sum_{i\in\cN_j}x^*_{ij}.
\end{align*}
The proof is completed by using the LP inequality that $\sum_{j\in G}\sum_{i\in\cN_j}x^*_{ij}/\sum_{j\in G}\lambda_j\ge s^*$.
\end{myproof}

\begin{myproof}[Proof of Theorem~\ref{thm:main-fa-ext2}]
$\serve(j)$ is fixed in advance, and hence independent from the number of arrivals of type $j$, for any $j\in J$.
Therefore, the expected number of an online type $j$ served is
\begin{align*}
\E[\min\{\Pois(\lambda_j),|\serve(j)|\}]
&=\E[\Pois(\lambda_j)\cdot\mathbf{1}(\Pois(\lambda_j)\le\lfloor x^*_j\rfloor)+|\serve(j)|\cdot\mathbf{1}(\Pois(\lambda_j)>\lfloor x^*_j\rfloor)] \\
&=\E[\Pois(\lambda_j)\cdot\mathbf{1}(\Pois(\lambda_j)\le\lfloor x^*_j\rfloor)]+\E[|\serve(j)|]\Pr[\Pois(\lambda_j)>\lfloor x^*_j\rfloor] \\
&=\E[\Pois(\lambda_j)\cdot\mathbf{1}(\Pois(\lambda_j)\le\lfloor x^*_j\rfloor)]+x^*_j\Pr[\Pois(\lambda_j)>\lfloor x^*_j\rfloor] \\
&=\E[\min\{\Pois(\lambda_j),x^*_j\}]
\end{align*}
where the first equality uses the property that $|\serve(j)|\in\{\lfloor x^*_j\rfloor,\lceil x^*_j\rceil\}$, the second equality uses independence, and the third equality uses the property that $\E[|\serve(j)|]=x^*_j$.

For any group $G\in\cG$, the expected fraction served is
\begin{align*}
\frac{\sum_{j\in G}\E[\min\{\Pois(\lambda_j),x^*_j\}]}{\sum_{j\in G}\lambda_j}
&\ge\frac{\sum_{j\in G}\frac{\E[\min\{\Pois(\lambda_j),\lambda_j\}]}{\lambda_j}\cdot x^*_j}{\sum_{j\in G}\lambda_j} \\
&\ge\frac{\E[\min\{\Pois(\lambda),\lambda\}]}{\lambda}\cdot\frac{\sum_{j\in G}x^*_j}{\sum_{j\in G}\lambda_j} \\
&=(1-\sfe^{-\lambda}\frac{\lambda^{\lambda}}{\lambda!})\cdot\frac{\sum_{j\in G}x^*_j}{\sum_{j\in G}\lambda_j}
\end{align*}
where the first inequality holds because $1\ge\frac{x^*_j}{\lambda_j}$,
and the second inequality holds because $\frac{\E[\min\{\Pois(\lambda),\lambda\}]}{\lambda}$ is increasing in $\lambda$.
Finally, $\frac{\sum_{j\in G}x^*_j}{\sum_{j\in G}\lambda_j}\ge s^*$ by LP feasibility, where $s^*$ is in turn an upper bound on $\OPT$.
Since this holds for all groups $G\in\cG$, the proof is complete.
\end{myproof}

\section{Missing Proofs from Section~\ref{sec:fb-small}}\label{app:fb}

\begin{myproof}[Proof of Theorem~\ref{thm:fb-small}.]
Let $b$ be the serving capacity of the single offline agent. Thus, the fairness of $\fcfs$ under $\fb$ should be at least $\Pr[\Pois(\lam) \le b]$. In contrast, the fairness of the offline optimal under $\fb$ should be $\OPT= \Pr[\Pois(\lam) \le b]+\sum_{k>b} \Pr[\Pois(\lam)=k] b/k$. By definition, the competitive ratio of $\fcfs$ under $\fb$ is at least 
\[
f(b,\lam) \doteq \frac{\Pr[\Pois(\lam) \le b]}{\Pr[\Pois(\lam) \le b]+\sum_{k>b} \Pr[\Pois(\lam)=k] b/k} .
\]

We now show that the value of $f(b,\lam)$, for all positive integers $b$ and $\lam\le1$, is lower-bounded by $f(1,1)$, which equals approximately 0.863. We first show that for any given $\lam  \le 1$, $f(b, \lam)$ is an increasing function of $b$ when $b \ge 1$. Fix a $\lam \in [0,1]$, Let $f_1(b)=\Pr[\Pois(\lam) \le b]$ and $f_2(b)=\Pr[\Pois(\lam) \le b]+\sum_{k>b} \Pr[\Pois(\lam)=k] b/k$. Thus, we have $f(b,\lam)=f_1(b,)/f_2(b)$. Observe that (1) $f_1(b)-f_1(b-1)= \Pr[\Pois(\lam)=b]=\sfe^{-\lam}\lam^b/b!$; (2) $f_2(b)-f_2(b-1)=\sum_{k=b}^{\infty} \sfe^{-\lam} \lam^k/(k \cdot k!)$. Thus, for $b\ge 2$, 
\[
\frac{f_1(b)-f_1(b-1)}{f_2(b)-f_2(b-1)}=\frac{\sfe^{-\lam}\lam^b/b!}{\sum_{k=b}^{\infty} \sfe^{-\lam} \lam^k/(k \cdot k!)} \ge 
\frac{1}{\sum_{k=b}^{\infty} b!/(k \cdot k!)} \ge \frac{1}{\sum_{k=2}^{\infty} 2!/(k \cdot k!)}=1.573.
\]
Note that $f_1(b)/f_2(b) \le 1$. Thus, we claim that $f(b,\lam) =f_1(b)/f_2(b)$ is an increasing function of $b \ge 1$. So, $f(b,\lam) \ge f(1,\lam)$. 

Now we show $f(1,\lam)$ is a decreasing function of $\lam \in [0,1]$. When $b=1$, we have
\[
f(1,\lam)=\frac{\sfe^{-\lam}(1+\lam)}{\sfe^{-\lam}(1+\lam)+\sum_{k=2}^{\infty} \frac{\sfe^{-\lam} \lam^k}{k! k}}=\frac{1}{1+\frac{\lam^k}{1+\lam}\sum_{k=2}^{\infty} \frac{ 1}{k! k}}.
\]
Observe that $\lam^k/(1+\lam)$ increases over $\lam>0$ for all given integer $k \ge 1$. Thus, we claim $f(1,\lam)$ is a decreasing function of $\lam$ over $\lam \in [0,1]$. Therefore, $f(1,\lam) \ge f(1,1) \sim 0.863$.
\end{myproof}

\begin{myproof}[Proof of Theorem~\ref{thm:hard-fb}.]
Consider such an instance $\cI(b,\lam)$ that $b=1$. Assume all online types are rare. In other words, with probability one, every online type has at most one arrival. For each $t \in [0,1]$, let $\sig(\lam,t)$ be the fairness achieved by an optimal online algorithm under $\fb$ when the online process is restricted as Poisson process of rate $\lam t$. Thus, we care about the value $\sig(\lam,1)$, which is the fairness achieved by the online optimal. 

Consider an infinitesimally small period $\del$ during which at most one arrival can occur. Now we try to upper bound $\sig(\lam,t+\del)$. (Case 1) There is no arrival during $(t,t+\del]$ which occurs with probability $\sfe^{-\lam \del}$. In the case, we have $\sig(\lam,t+\del)=\sig(\lam,t)$. (Case 2) There is one arrival during $(t,t+\del]$ which occurs with probability $1-\sfe^{-\lam \del}$. In this case, we have $\sig(\lam,t+\del) \le \min(\sig(\lam,t), 1-\sig(\lam,t)+\sfe^{-\lam t})$, which is shown as below.

Let $\alp_{t,k}$ be the fairness achieved by an online optimal when there are $k$ arrivals during $[0,t]$. Observe that $\alp_{t,0}=1$ for all $t \in [0,1]$. Therefore, by definition, $\sig(\lam,t)=\sum_{k=0}^{\infty} \alp_{t,k} \Pr[\Pois(\lam t)=k]$. Assume there is one arrival during $(t,t+\del]$. Note that
\begin{align*}
\sig(\lam,t+\del) &=\sum_{k=0}^\infty \min( \alp_{t,k}, 1-k \cdot \alp_{t,k})\Pr[\Pois(\lam t)=k] \le \sum_{k=0}^\infty \alp_{t,k}\Pr[\Pois(\lam t)=k]=\sig(\lam,t), \\
\sig(\lam,t+\del) & \le \sum_{k=0}^\infty ( 1-k \cdot \alp_{t,k})\Pr[\Pois(\lam t)=k]\le
1- \sum_{k=1}^\infty\alp_{t,k}  \Pr[\Pois(\lam t)=k] = 1-(\sig(\lam,t)-\sfe^{-\lam t}).
\end{align*}
Thus, we claim that $\sig(\lam, t+\del) \le \min(\sig(\lam,t), 1-\sig(\lam,t)+\sfe^{-\lam t})$. Wrapping up all the above analysis, we have $\sig(\lam,t+\del) \le \sfe^{-\lam \del} \sig(\lam,t)+(1- \sfe^{-\lam \del})\min(\sig(\lam,t), 1-\sig(\lam,t)+\sfe^{-\lam t})$. This suggests that $\partial \sig(\lam,t)/\partial t \le -\lam \sig(\lam,t)+\lam \min(\sig(\lam,t), 1-\sig(\lam,t)+\sfe^{-\lam t})$. 

For each given $\lam \in [0,1]$, let $R_\lam(t)$ be the unique function satisfying that $dR_\lam(t)/dt=-\lam R_\lam(t)+\lam \min(R_\lam(t), 1-R_\lam(t)+\sfe^{-\lam t})$ with $R_\lam(0)=1$. Thus, we claim that $\sig(\lam,1) \le R_\lam(1)$. Recall that the offline optimal has a performance of 
$\sfe^{-\lam}(1+\lam)+\sum_{k=2}^{\infty} \frac{\sfe^{-\lam} \lam^k}{k! k}$ under $\fb$. We can numerically verify that $R_\lam(1)/\big(\sfe^{-\lam}(1+\lam)+\sum_{k=2}^{\infty} \frac{\sfe^{-\lam} \lam^k}{k! k}\big)$ gets its minimum value of $0.942$ when $\lam=1$. Thus, we establish our result. 
\end{myproof}

\begin{myproof}[Proof of Lemma~\ref{lem:off-fb-ext}.]
Consider an arrival vector  $\cA$ and let $A=\sum_j A_j$ be the total arrivals of all online agents. Observe that (1) $\eta(\cA)=1$ when $A \le b$; (2) $\eta(\cA)=b/k$ when $A=k>b$. Therefore,
\begin{align*}
\off&=\E_{\cA}[\eta(\cA)]=\Pr[A \le b] \cdot 1+\sum_{k>b}^{\infty}\Pr[A=k] \cdot \frac{b}{k} \le   \Pr[A \le \lam (1-(1-\kap))]+ b \cdot \sum_{k=1}^\infty \frac{\sfe^{-\lam}\lam^k}{k!}\frac{1}{k} 
\\
& \le\exp\bp{-\frac{\lam(1-\kap)^2 }{2}}+b \cdot \bp{\frac{1}{\lam}+\frac{1}{\lam^2}+o\bsp{\frac{1}{\lam^2}}}=\kap\bp{1+\frac{1}{\lam}+o\bsp{\frac{1}{\lam}}}.
\end{align*}
Note that the inequality on the last line is due to the lower tail bound of a Poisson random variable as shown by~\cite{pois-tail}. Another trick involved is 
$ \sum_{k=1}^\infty \frac{\lam^k}{k!}\frac{1}{k}=\mathrm{Ei}(\lam) - \ln \lam - \gam$ where  $\gam \doteq \lim_{n \rightarrow \infty} \bp{\sum_{k=1}^n 1/k - \ln n}\sim 0.577$ is a constant, and $\mathrm{Ei}$ is the Exponential integral function. As shown by~\cite{masina2019useful}, $\mathrm{Ei}(\lam)=(\sfe^\lam/\lam) (1+1/\lam+o(1/\lam))$ when $\lam \gg 1$. Thus, we are done.
\end{myproof}

\begin{myproof}[Proof of Theorem~\ref{thm:fb-large}.]
For notation convenience, we use $\cI$ to denote $\cI(b,\lam)$ and $\ALG$ to denote the probabilistic-rejection algorithm. By definition, we have $\ALG(\cI)=\E_{\cA} \bb{\min_{j: A_j>0} \E_{\ALG}[X_j]/A_j} \doteq \E_{\cA} [\ALG(\cA)]$. Consider a given arrival vector $\cA$ with $A$ being the total number of online arrivals. By definition, we have $\ALG(\cA)=1$ when $A=0$. 

Now we show that  $\ALG(\cA) = b/K$ conditioning on (1) $0<A \le K$ and (2) $b/K \le 1$. Note that by dependent rounding, we have (P1) $\Pr[Y_j=1]=b/K$ for all $j \in [K]$ and (P2) $\Pr\bsb{\sum_{j=1}^K Y_j \le \sum_{j=1}^K b/K=b}=1$. Focus on a given $j$ with $A_j>0$. Consider a specific online arrival of type $j$, which is counted as the $k$th  arrival among all types of online agents. When $A \le K$, we see that $k \le K$ and the single offline agent will not reach the capacity upon the arrival due to (P2). Thus, we claim that the type-$j$ agent will be served with probability equal to $\Pr[Y_k=1]=b/K$ for each of its $A_j$ arrivals. Thus, $\E[X_j]=A_j \cdot b/K$ and $\ALG(\cA)=\min_{j: A_j>0} \E_{\ALG}[X_j]/A_j =b/K$. Consider the following three cases.

(Case 1) $b>\lam$. In this case, $K=b$. If $A=0$, $\ALG(\cA)=1$ and if $0<A \le K$, $\ALG(\cA) = b/K=1$. Thus, we claim that $\ALG (\cA)=1$ when $A \le K$.
\[
\ALG(\cI)= \E_{\cA}[\ALG(\cA)]\ge \Pr[A \le K] =1-\Pr[\Pois(\lam) >b] \ge 
1-\exp\big(-\lam (\kap-1)^2/(2 \kap)\big).
\]
Note that $\off(\cI) \le 1$. Thus, $\ALG(\cI)/\off(\cI) \ge \ALG(\cI)$ and we are done. 
\\
(Case 2)  $b=\lam$. In this case, when $A \le K$, $\ALG(\cA) \ge b/K \ge 1/(1+\ep)$. Thus,
\[
\ALG(\cI)/\off(\cI)  \ge \ALG(\cI)=\E_{\cA}[\ALG(\cA)]\ge   \frac{\Pr[A \le K]}{1+\ep} \ge 
\bp{1- \exp\bp{-\frac{\lam \ep^2}{2(1+\ep)}}} \cdot \frac{1}{1+\ep}. 
\]
By taking $\ep=\sqrt{\ln \lam/\lam}$, we establish our claim. \\
(Case 3)  $b<\lam$: we have $\ALG(\cA) \ge b/K \ge \kap/(1+\ep)$ when $A \le K$. 
Thus,
\[
\ALG(\cI)=\E_{\cA}[\ALG(\cA)]\ge \Pr[A \le K]  \frac{\kap}{1+\ep} \ge 
\bp{1- \exp\bp{-\frac{\lam \ep^2}{2(1+\ep)}}} \cdot \frac{\kap}{1+\ep}. 
\]

By Lemma~\ref{lem:off-fb-ext}, we have that for any given instance $\cI(b,\lam)$ with $b<\lam$,
\[
\frac{\ALG(\cI)}{\off(\cI)} \ge \bp{1- \exp\bp{-\frac{\lam \ep^2}{2(1+\ep)}}} \cdot \frac{1}{1+\ep} \cdot \frac{1}{1+1/\lam+o(1/\lam)}.
\]
By taking $\ep=\sqrt{\ln \lam/\lam}$, we establish our claim. 
\end{myproof}

\end{APPENDICES}

\end{document}